# Science at Risk: The Urgent Need for Institutional Support of Long-Term Ecological and Evolutionary Research in an Era of Data Manipulation and Disinformation

Steering committee for CNRS long-term studies in Ecology and Evolution (SEE-Life)

Vincent A Viblanc[1], Élise Huchard[2], Gilles Pinay[1], Elena Ormeno[3], Céline Teplitsky[4], François Criscuolo[1], Dominique Joly[1], David Renault[5], Cécile Callou[1], Françoise Gourmelon[1], Sandrine Anquetin[6], Bénédicte Augeard[7], Fabienne Aujard[1], Sophie Ayrault[6], Philippe Grandcolas[1], Agathe Euzen[1], Agnès Mignot[1*] & Stéphane Blanc[1*]

[1]CNRS Écologie & Environnement, CNRS, 3 rue Michel-Ange 75794 - PARIS Cedex 16, France
[2]Institut des Sciences de l'Évolution de Montpellier (ISEM), Université de Montpellier, CNRS UMR 5554, IRD, EPHE, Montpellier, France
[3]CNRS UMR 7263, Université Aix-Marseille, Université d'Avignon, IRD, IMBE, Marseille, France
[4]CEFE, Univ Montpellier, CNRS, EPHE, IRD, Montpellier, France
[5]Institut Polaire Français Paul Émile Victor (IPEV), Technopôle Brest Iroise, CS 60075, 29280 Plouzané, France
[6]CNRS Terre & Univers, CNRS, 3 rue Michel-Ange 75794 - PARIS Cedex 16, France
[7]Office Français de la Biodiversité

Contributions: This paper was written by the direction of the CNRS Écologie & Environment and the Steering Committee for the CNRS Long-Term Studies in Ecology and Evolution (SEE-Life) program. It was read, and endorsed by the following list of persons involved in SEE-Life studies:

Co-Signatories (endorsements of the contents of this paper):

- Christophe Barbraud; Senior Researcher; Centre d'Etudes Biologiques de Chizé, UMR 7372, CNRS – La Rochelle Université
- Aurélien Besnard; Senior Researcher; CEFE, Univ Montpellier, CNRS, EPHE, IRD, Montpellier, France
- Clotilde Biard; Lecturer; Sorbonne Université, Université Paris Cité, Univ Paris Est Créteil, CNRS, IRD, INRAE, Institut d'écologie et des sciences de l'environnement de Paris, IEES, F-75005 Paris, France
- Christophe Bonenfant; Researcher; Laboratoire de Biométrie et Biologie Evolutive UMR 5558, CNRS, Université Claude Bernard Lyon 1, Villeurbanne, France.
- Thierry Boulinier; Senior Researcher; CEFE, Univ Montpellier, CNRS, EPHE, IRD, Montpellier, France

- Vincent Bretagnolle; Senior Researcher; Centre d'Etudes Biologiques de Chizé, UMR 7372 CNRS – La Rochelle Université – Zone Atelier Plaine & Val de Sèvre
- Emmanuelle Cam; Professor; Laboratoire LEMAR, UMR 6539, Université de Bretagne Occidentale (UBO), CNRS, IRD, IFREMER, Institut Universitaire Européen de la Mer (IUEM), Place Nicolas Copernic, 29280 Plouzané, France
- Simon Chamaillé-Jammes; Senior Researcher; CEFE, Univ Montpellier, CNRS, EPHE, IRD, Montpellier, France
- Anne Charmantier; Senior Researcher; CEFE, Univ Montpellier, CNRS, EPHE, IRD, Montpellier, France
- Marie JE Charpentier; Senior Researcher; Institute of Evolutionary Sciences of Montpellier (ISEM), Université de Montpellier, CNRS, IRD, EPHE, Montpellier, France
- Julien Cucherousset; Senior Researcher; Centre de Recherche sur la Biodiversité et l'Environnement (CRBE), Université de Toulouse, CNRS, IRD, Toulouse INP, Université Toulouse 3 – Paul Sabatier (UT3), Toulouse, France 2 FRB — CESAB, Montpellier, France
- Pierre de Villemereuil; Lecturer; Institut de Systématique, Évolution, Biodiversité (ISYEB, UMR 7205), École Pratique des Hautes Études, PSL University, MNHN, CNRS, SU, UA, Paris, France.
- Claire Doutrelant; Senior Researcher; CEFE, Univ Montpellier, CNRS, EPHE, IRD, Montpellier, France
- Julie Duboscq; Researcher; Eco-Anthropology Lab (UMR7206, CNRS-MNHN-Université Paris Cité)
- Christine Dupuy; Professor; Littoral, Environnement et Sociétés (LIENSs), UMR 7266 CNRS - La Rochelle Université, La Rochelle, France
- Olivier Duriez; Lecturer; CEFE, Univ Montpellier, CNRS, EPHE, IRD, Montpellier, France
- Bruno Faivre; Professor; Université Bourgogne Europe, UMR EPHE CNRS Biogéosciences
- Jérôme Fort; Senior Researcher; Littoral, Environnement et Sociétés (LIENSs), UMR 7266 CNRS - La Rochelle Université, La Rochelle, France
- Sabrina Gaba; Senior Researcher; USC1339 CNRS INRAE La Rochelle Université Centre d'Etudes Biologiques de Chizé
- Marlène Gamelon; Researcher; Laboratoire de Biométrie et Biologie Evolutive UMR 5558, CNRS, Université Claude Bernard Lyon 1, Villeurbanne, France.
- Jean-Yves Georges; Senior Researcher; Université de Strasbourg, CNRS, IPHC UMR7178, F-67000 Strasbourg, France
- David Grémillet; Senior Researcher; CEFE, Univ Montpellier, CNRS, EPHE, IRD, Montpellier, France
- Pierre-Yves Henry; Professor; Centre de Recherches sur la Biologie des Populations d'Oiseaux (CRBPO), Centre d'Ecologie et des Sciences de la Conservation (CESCO UMR 7204 MNHN CNRS SU), Mécanismes adaptatifs et évolution (MECADEV UMR 7179 MNHN CNRS), PatriNat (UAR 2006 MNHN OFB CNRS IRD)
- Eric Imbert; Lecturer; Institute of Evolutionary Sciences of Montpellier (ISEM), Université de Montpellier, CNRS, IRD, EPHE, Montpellier, France
- Philippe Jarne; Senior Researcher; CEFE, Univ Montpellier, CNRS, EPHE, IRD, Montpellier, France
- Akiko Kato; Research Engineer; Centre d'Etudes Biologiques de Chizé, UMR 7372, CNRS – La

- Rochelle Université
- Céline Le Bohec; Senior Researcher; CEFE, Univ Montpellier, CNRS, EPHE, IRD, Montpellier, France; Centre Scientifique de Monaco, Département de Biologie Polaire, Monaco, Principality of Monaco
- Jean-François Lemaître; Senior Researcher; Laboratoire de Biométrie et Biologie Evolutive UMR 5558, CNRS, Université Claude Bernard Lyon 1, Villeurbanne, France.
- Jean-Marc Limousin; Researcher; CEFE, Univ Montpellier, CNRS, EPHE, IRD, Montpellier, France
- Romain Lorrilliere; Research Engineer; Centre de Recherches sur la Biologie des Populations d'Oiseaux (CRBPO), Centre d'Ecologie et des Sciences de la Conservation (CESCO UMR 7204 MNHN CNRS SU), Muséum national d'Histoire naturelle, Centre National de la Recherche Scientifique, Sorbonne Université, CP 135, 57 rue Cuvier 75005 Paris, France
- Sylvie Massemin; Professor; Université de Strasbourg, CNRS, IPHC UMR7178, F-67000 Strasbourg, France
- Jean-Philippe Mevy; Lecturer; CNRS UMR 7263, Université Aix-Marseille, Université d'Avignon, IRD, IMBE, Marseille, France
- Victor Narat; Researcher; Eco-Anthropology Lab (UMR7206, CNRS-MNHN-Université Paris Cité)
- Christophe Petit; Lecturer; Institute of Evolutionary Sciences of Montpellier (ISEM), Université de Montpellier, CNRS, IRD, EPHE, Montpellier, France
- Benoit Pujol; Researcher; PSL Université Paris: EPHE-UPVD-CNRS, UAR 3278 CRIOBE, Université de Perpignan, 52 Avenue Paul Alduy, CEDEX 9, 66860 Perpignan, France
- Francis Raoul; Professor; Université Marie et Louis Pasteur
- Jean-Patrice Robin; Senior Researcher, retired; Université de Strasbourg, CNRS, IPHC UMR7178, F-67000 Strasbourg, France
- Carsten Schradin; Senior Researcher; Université de Strasbourg, CNRS, IPHC UMR7178, F-67000 Strasbourg, France
- Jerome Spitz; Researcher; Pelagis, UAR3462 CNRS - La Rochelle Université
- Antoine Stier; Researcher; Université de Strasbourg, CNRS, IPHC UMR 7178, F-67000 Strasbourg, France
- Christophe Thébaud; Distinguished professor; Centre de Recherche sur la Biodiversité et l'Environnement, UMR 5300 CNRS-UT-IRD-TINP

Correspondence: cnrs-ecologie-seelife@cnrs.fr

# Abstract

Planet Earth and the biodiversity it supports are in crisis. Human impact on terrestrial, marine and freshwater ecosystems and the hundreds of thousands of organisms that inhabit them is global. To what extent can we push ecosystems before they collapse? Will species adapt to these changes and at what rate? What are the consequences, for the environment and humankind? These are some of the most pressing issues to date. Clear answers can only be addressed through long-term research programs that are extremely complex in their deployment, and by the analyses of the unique data they produce on species and ecosystem responses to change. Yet, too little institutional support and consideration have been given to long-term ecological and evolutionary research.

We describe the action recently taken by the *French National Center for Scientific Research (CNRS)* to recognize and support long-term ecological and evolutionary research. We provide some salient examples of critical knowledge attainable only through long-term studies in ecology and evolution, before highlighting how global institutional schemes can not only support long-term research, but lead to informed conservation efforts and societal change. Now more than ever, as populism grows and fuels mis- and dis-informed politics, governmental programs are urgently needed to support data collection, establish data-grounded facts, inform political spheres, and refuel trust with society at large.

Keywords: conservation, ecology, environmental policy, evolution, long-term research

# Responding to biodiversity loss in times of political turmoil: the need for national research infrastructures for ecology and evolution

In early 2025, several leading environmental datasets maintained by national agencies in countries recently marked by populist electoral shifts were abruptly taken offline or replaced with curated versions that obscure or distort previously accessible information. These actions, framed as "streamlining," in fact erased decades of scientific work and undermined the empirical foundations of environmental policy. Even more troubling, multiple media investigations have confirmed the political orchestration behind these decisions, with data being actively revised to downplay trends in *biodiversity loss*, *climate change*, *pollution and human responsibility*. This alarming erosion of transparency transforms once-robust scientific knowledge into misleading or outright false narratives. It sets a dangerous precedent: when long-term data becomes a target, our ability to understand — and respond to — global environmental change is profoundly compromised.

At the same time, the [World Economic Forum (WEF) Global Risks Report 2025](#) identified "misinformation and disinformation" as the top short-term threat to global stability, surpassing geopolitical conflicts and economic uncertainty. In science, this threat materializes through underfunded institutions, politicized research agendas, and discredited experts[1,2] — all in a context where public trust is rapidly eroding[3,4].

Now more than ever, long-term ecological and evolutionary research must be recognized as essential infrastructure, akin to national health systems or weather services. These studies are the only way to detect and understand slow, cumulative processes — such as biodiversity loss, species adaptation, or ecosystem collapse — that are invisible in short-term snapshots. Yet in many countries, these efforts are not systematically supported, and even in well-resourced research systems, they often depend on short-term grants or precarious labor.

In France, the National Center for Scientific Research (CNRS), France's largest public research institution (more than 1,100 laboratories, over 33,000 personnel including 28,000 scientists), has taken steps to change this by creating the SEE-LIFE **program** ([https://www.inee.cnrs.fr/fr/see-life-suivis-long-terme-en-ecologie-et-evolution)](https://www.inee.cnrs.fr/fr/see-life-suivis-long-terme-en-ecologie-et-evolution), a national framework to recognize, support, and coordinate long-term studies in ecology and evolution. This initiative is not just a scientific milestone — it is a political one. It affirms that public knowledge, grounded in sustained observation and shared openly, is a democratic good worth protecting. It is a model that can and should be replicated internationally.

As we face increasingly sophisticated attempts to manipulate facts and discredit science, long-term ecological and evolutionary research — and the institutions that sustain it — are not luxuries. They are our best defense.

## Why are long-term studies of ecosystems and biodiversity so important?

We know that biodiversity plays a key role in maintaining the function and stability[5] of the food webs and ecosystems on which humanity depends. Understanding its resilience and ability to respond to global change requires long-term data[6–8].

**First**, long-term studies of ecology and evolution provide the tool for assessing how changes occurs over time at all levels of organization (from organisms to ecosystems). Short-term studies, often important in decrypting underlying processes and mechanisms at the individual level, only provide partial and fragmentary information on the variability of conditions occurring in natural ecosystems over time and on organisms' response to environmental change. For instance, results on population demographics from short-term field studies do not always scale-up to those found over decades[9], making it hard to predict the outcome of populations in the face of global change. Meta-analyses indicate that defining population trends require studies spanning a minimum of 10 years[10,11] (sometimes up to 20 years[10]) to generate reliable results. Therefore, conclusions about conservation measures or changes in human practices delivered to decision-makers based on short-term studies may turn out to be partly misguided over the longer term, fueling science skepticism from the public. **Second**, long-term studies allow deciphering the mechanisms through which global changes affect populations, species and communities, and through which adaptation occurs[12,13]. Understanding these mechanisms (*e.g.*, genetic, epigenetic, physiological, behavioral) improves our understanding of how evolution operates and how variable responses may be, in turn leading to more informed conservation actions[14]. **Third**, long-term studies allow understanding not only how organisms adapt to environmental change, but how such change might feedback on the environment both on short and longer timescales. In essence, they allow reconciling slow (where changes in the environment trigger evolution) and fast (where rapid evolution affects ecology) eco-evolutionary processes[15,16]. In particular, long-term studies that concurrently monitor environmental parameters (e.g., temperature, pH, salinity) together with species population rates and phenotypic traits permit identifying the factors that modulate ecological and evolutionary rates simultaneously, shaping eco-evo dynamics[16]. Although meta-analyses based on short-term databases are increasingly used as promising statistical approaches to assess the generality of evolutionary processes giving rise to macroecological patterns[17], they are not a substitute for long-term studies. **Finally**, long-term studies, beyond their scientific output, play a critical role in inclusive education and structuring of the scientific community by training students, transmitting skills and knowledge, and more generally shaping the scientific agenda and the whole disciplinary culture[18]. This is especially true in many Southern countries where some long-term studies run, and where opportunities for training are scarcer than in the northern hemisphere. Long-term studies also constitute a unique exchange forum among professionals originating from different horizons, as these are often conducted with non-academic partners, and/or include educational objectives or bilateral exchanges of knowledge on ecosystems with local communities.

As natural ecosystems and their associated biodiversity are key to solving the climate crisis by providing realistic, sustainable and costless solutions to mitigate carbon emissions[19,20]; as the importance of healthy ecosystems for humanity is underlined by the astronomical economic costs of services they provide (e.g. pollination, photosynthesis, carbon sequestration, water regulation, control of pest damages and emerging zoonoses, resilience against climatic events, etc., estimated at some $125 (USD) trillion per year[21]); and as we have a moral and ethical obligation towards the environment we share with other species; the importance of long-term ecological and evolutionary research cannot – and should not – be understated.

## Three salient examples

### Long-term studies provide the fundamental knowledge necessary for informed forecasting

A common specificity of long-term studies is to document the life cycles, life history traits, ecological characteristics and geographical distribution of generations of organisms over extended periods of time[12]. Understanding how and why individuals of different species use habitats and resources the way they do constitutes the fundamental knowledge that allows us to understand how the living world responds to change and, arguably more importantly, predict changes to come. For instance, under globalization, invasive alien species are increasing worldwide with no tangible evidence that the rate of biological invasions is slowing down[22,23] (**Fig. 1, left**). Their environmental and economic cost is massive (estimated at $1,288 trillion (USD, 2017 value) over 1970-2017, and $162,7 billion in 2017 alone) and grossly underestimated[24]. The raw damage cost for biological invasions from 1980-2019 worldwide ($1,208.0 billion USD, 2020 value) is akin to that of natural hazards[25], second most important to storms ($1,913.6 bn), higher than earthquakes ($1,139.4 bn) or floods ($1,120.2 bn), and an order of magnitude above wildfires ($138.2 bn). Understandably, the development of tools allowing us to forecast the invasive potential of species in a rapidly changing environment are urgently needed. Tailoring such tools uniquely relies on our knowledge of species ecology and life cycle characteristics[26], much of which comes from long-term studies of organisms and ecosystems. In

addition to forecasting, long-term studies critically allow evaluating: (1) the actual dynamic of invasions, currently suffering from poor-updating[27], (2) the evolutionary responses of native species to biological invasions[28], (3) the effectiveness and economic costs of human intervention in controlling biological invasions (e.g., spectacular recovery of seabird populations following rat eradications on islands[29–31]), and (4) the direct or indirect impact of invasive species on human quality of life[32].

Long-term studies allow detecting early warning signals of ecosystem shifts and population collapse

Predicting the tipping points at which major changes in ecosystem or population dynamics occur is notoriously complex. Over the past two decades, research has theorized and tested the existence of early warning statistical signals in long-term data series (such as changes in spatial or temporal autocorrelation or increases in variance) that may allow forecasting the point in time when critical shifts occur[33,34,34–39]. Including trait-based measures of change (such as changes in species size distribution) in addition to changes in abundance metrics, substantially improves the predictive power of early warning signals[40–42], and is of particular importance to management-decision-making. A good example concerns the global collapse of iconic whale populations during the 20$^{th}$ century[43] (**Fig. 1, middle**). Analyses of changes in whale body size together with historic data on catches suggest early warning signals were detectable up to 40 years – some 4 sperm-whale generations – before the global collapse of whale populations occurred[42]. Reaching beyond classically-measured traits such as body size or mass, long-term studies of physiological, molecular or genetic components of organismal biology may allow profound insights into early warning signals. For instance, the advent of molecular biology techniques has recently allowed us to probe into the state of specific molecular DNA sequences indicative of cellular and organism aging, and to find that such information allows determining the age-structure of a population, and forecasting the decline and extinction of animal populations in the wild[44,45]. Further insights into early warning signals are likely to unfold as omics methodologies develop and become increasingly cheaper and used[46]. This example highlights the power of long-term trait-based, physiological and mechanistic monitoring designs in predicting population or ecosystem outcomes in the future. Critically, research suggests that the predictive power of early warning signals decreases with decreasing length of the time series[47], emphasizing the need for long-term monitoring.

Long term studies allow understanding eco-evo dynamics providing the data to establish sustainable economic activities

The intertwined nature of ecology and evolutionary processes has been known as far back as Darwin[48] and Fisher[49]. Despite this, true studies of eco-evo dynamics that consider rapid evolution within contemporary time-scales or ecological feedbacks caused by evolutionary change are still thin on the ground[15,50–52]. Yet, such studies are desperately needed if one wants to predict how organisms and ecological environments respond to global change[53]. Consider the case of commercial fishing, an important source of revenue in the global economy (estimated at $157 bn[54], USD 2022 value), yet with massive consequences for biodiversity. From a biodiversity perspective, one of the main issues of commercial fisheries is the bycatch of unwanted species from invertebrates and fish, to marine mammals, seabirds and sea turtles[55] (**Fig. 1, right**). For iconic albatrosses for instance, fishery bycatch mortality is a major global cause of population decline[56,57]. This has led to the suggestion that bycatch may act as a strong selection pressure against attraction to boats[58], facilitating the evolutionary rescue of critically endangered species from extinction[59]. Yet, albatross attraction to fishing vessels varies with age, sex, and is reinforced in bolder individuals. Since boldness is a heritable personality trait, and the fitness benefits/costs of fishery discards differ between personality types, sexes, and age classes, complex eco-evo feedbacks are likely to lead to differential selection for or against vessel attraction in albatross populations[60]. Such knowledge would be impossible without long-term individual-based studies and critically shows that predicting the response of wild animal populations to human activities and developing adequate management practices requires accounting for complex eco-evo dynamics. As a whole, the longer the time series is, the better scientific theories can be tested and dynamics understood.

## The role and responsibility of institutional support

Despite the incredible wealth of knowledge produced by long-term ecological and evolutionary research[61,62], including a disproportionate contribution to shaping environmental policy[63], the chief threats looming over long-term studies is an overall lack of recurring funding schemes[63–65] and institutional support and visibility[65,66].

Recognizing this, the CNRS SEE-LIFE program aims to support long-term Studies in Ecology and Evolution led by research teams of CNRS Ecology & Environment by providing: (1) an institutional endorsement, and (2) recurrent financial support for long-term ecological and evolutionary research, across all biomes, phylum and kingdoms. In addition, it (3) recognizes long-term studies in ecology and evolution as a research infrastructure in its own right, playing a key role in generating actionable data for all stakeholders. The SEE-LIFE endorsement provides national and international visibility to long-term ecological and evolutionary projects and runs parallel to the few (but substantive) global initiatives such as the U.S. National Science Foundation Long-Term Ecological Research (LTER) network supporting 27 LTER sites, or the integrated pan-European Long-Term Ecosystem, critical zone and socio-ecological Research (eLTER) network hosting over 500 sites across the world. In contrast to such global initiatives that focus on ecosystem-level function and processes, long-term SEE-LIFE studies originally span individual- through to population-, community-, and ecosystem-levels, addressing questions on ecological and evolutionary dynamics both from a fundamental and applied perspective. One common specificity across SEE-LIFE studies is that they go beyond the sole monitoring of ecosystem and population health through count estimates of population abundance or species presence/absence, to address the causal genetic, phenotypic (behavioral, physiological and morphological) and populational (interactions, migration) mechanisms underpinning organism capacity for change. SEE-LIFE studies thus aim at reinforcing eco-evo dialogues to understand contemporary evolution in the era of global change.

At its initial launch in 2023, the CNRS selected 64 projects (several of which had multiple long-term monitoring field sites) and committed to fund these projects over a 5-year period, renewable over time. Collectively, these studies span 23 French research centers, over 100 international collaborators, who monitor more than 500 species for over 10 to 102 year-long periods, and cover the vast majority of biomes on earth (**Fig. 2**). Together, they have already produced over 3000 publications, 500 reports in the grey literature, and contributed to the training of over 4900 personnel, including over 800 PhD students and post-doctoral fellows. Several of these studies are further incorporated into interdisciplinary and global networks such as the eLTER discussed above – and many have directly contributed to informing policy. Their contribution to our knowledge on ecosystems, fundamental ecology, evolution and conservation is therefore tremendous.

Based on SEE-LIFE objectives, we identify *key areas* where institutions can play pivotal roles in supporting long-term studies, and – critically – transferring knowledge into actionable policies (**Fig. 3**).

**Providing baseline recurrent funding and building a community**

The costs involved in maintaining long-term ecological and evolutionary research programs are numerous and include the costs of data collection (personnel, permits, logistics, equipment, consumables) and curation (personnel, storage), data analyses (personnel, laboratory equipment and consumables, storage and information technologies), and data valorization (personnel, publications, reports, books, etc.). For some biomes and studies, the logistics themselves are tremendous. Long-term studies of polar ecosystems in one of the most remote places on Earth, for instance, require some 1.85 M€ per year for logistic infrastructure and gross salary of the staff from the French Polar Institute resource agency alone (source: French Polar Institute Paul Emile Victor). What is fundamental to understand is that the collection of ecological data requires personnel in the field, often over prolonged periods of time. Data comes from observing, trapping, marking, sampling, releasing, and overall monitoring animals and ecosystems. Here lies the crux of the matter: the financial cost of perennially staffing all long-term research studies is substantial and often prohibitive for public institutions. In this context, how can we proceed?

The first simplistic – realistic and rapid – answer is: by *providing a minimal baseline amount*. The annual funding provided to individual SEE-LIFE studies is capped at a 10k€ maximum per study (range 3 to 10k€, average 7k€). Whereas this does not solve the issue of additional human resources (besides principal investigators whose salary is often covered by the institution), it does allow a buffer during lean periods of funding. Baseline support is not subject to objective-oriented priorities defined by political agendas, other than the continuation of long-term data collection.

The second – more complex and longer-term – solution is: by *building a community* that together may not only network towards mutual aid, exchange of expertise and workforce, but together and with institutional support may advocate and tip the scales towards finding more perennial solutions to the issue of human resources. Our experience of building such a community by launching the 1st SEE-LIFE Conference in Paris, 9-12 Sept 2024 (https://see-life-cnrs2024.sciencesconf.org) has provided some valuable lessons. First, feedback was overwhelmingly positive. Individual researchers were in need of a forum where long-term studies (their potential, pitfalls and requirements) could be discussed. Second, despite a generally good knowledge of the work of colleagues, many researchers were made aware of teams they did not know, working on similar questions and facing similar constraints than those they experienced. This provided an opportunity for rich exchanges on issues but also potential solutions, working at a national level.

The third, is that building an inclusive community allows the institution to better identify critical needs at a national scale, providing arguments to defend scientific interests in an overall national strategy, prioritizing short- to mid- to long-term actions. For instance, having recognized the predicament of university personnel involved in long-term studies that must conciliate teaching duties with fieldwork, the CNRS Ecology & Environment is currently working on finding solutions to alleviate the workload and support these categories of personnel.

**Providing key infrastructures and logistic resources**

Another critical role for institutions in supporting long-term research in ecology and evolution is by *providing key infrastructure and logistics* to support long-term data collection. Whereas long-term studies must be considered as *in natura* research infrastructures in their own right, institutions may reinforce the value of long-term projects by facilitating connections with other infrastructures or existing major equipment that transcend disciplinary communities (**Fig. 4**). For instance, biological field stations[67] (https://geo.igb-berlin.de/maps/169), experimental ecological platforms (e.g., https://www.anaee-france.fr/en/), or observatories focused on the geophysical components of the environment (e.g., https://www.insu.cnrs.fr/en/observatories-sciences-universe-osu), are of high added-value in supporting/complementing long-term studies for a holistic understanding of species and ecosystem responses to change. A key role for scientific institutions is then to identify opportunities between research infrastructures and promote interdisciplinary research through the organization of workshops and specific funding calls, between scientific communities ranging from the humanities to social sciences, mathematics, chemistry, physics, biology, paleontology, bioarcheology and ecology, federating researchers around related questions and methodologies (**Fig. 4**). For instance, networking paleontological, environmental archeology, molecular biology and ecological communities may facilitate bridging macro and micro-evolutionary processes[68].

Institutions can also play a key role in supporting long-term research projects through data infrastructures designed to curate, store and share long-term data, abiding by FAIR (Findable, Accessible, Interoperable, Reusable) principles[69]. For instance, SEE-LIFE projects are built back-to-back on the DoHNEE (previously BBEES; https://bbees.mnhn.fr/en) research infrastructure tasked with guiding researchers in the standardization and storage of collected data in a secure data repository (https://data.indores.fr/). Such infrastructures not only allow critically safeguarding and sharing long-term data and protocols across successive generations of researchers, they provide a tool *for large-scale comparative analyses* and a repertoire allowing the transfer of actionable data to public policies[70]. In essence, they act as repositories for a scientific heritage on the biosphere.

**Coordinating across agencies and reaching the political sphere**

The building of a community under the umbrella of an institutional program allows *reaching beyond the academic sphere to stakeholders and decision-makers.* This, arguably, is where institutional support has the most to bring, both by advocating on the importance of long-term studies at a governmental level, and by facilitating the inclusion of individual scientists into working groups that enact policy, both nationally and internationally. Institutions often have a broad view of governmental priorities and working bodies, allowing for the *coordination of joint-programs across state levels* that are inclusive of long-term studies. As an example, the first framework agreement aimed at reinforcing and coordinating research on the biodiversity conservation was signed between the CNRS and the French Office for Biodiversity (OFB) in September 2024. This was followed by a round table during the 1$^{st}$ SEE-LIFE Conference involving members of the CNRS, the OFB, the French Ministry for Education and Research, the French Ministry for Ecological Transition and territory cohesion, the Foundation for Research on Biodiversity, and various French Experts – leading to a national recognition of the importance of long-term studies in ecology and evolution in understanding biodiversity responses to change. Another example is in coordinating with the French Polar Institute (https://institut-

polaire.fr/en/) to support long-term field research in the remotest places on Earth. By co-jointly identifying key long-term projects, substantial logistic and financial means provided by the French Polar Institute (a public interest group) can be tailored and directed to specific research programs more efficiently, avoiding redundant and time-consuming applications (for individual researchers) and evaluations (for reviewers and committees). This type of ongoing interaction between institutions running interrelated research programs further facilitates rapid scientific logistic and human-resource responses to exceptional environmental events, such as that being currently coordinated between the French Polar Institute and the CNRS on the 2024-25 avian H5N1 flu outbreak spreading across the sub-Antarctic (emergency 190 k€ fund invested by the CNRS Ecology & Environment and CNRS Biology in 2025). Such coordinated efforts will provide critical scientific data on transmission phenomena within and across biological reservoirs, to understand the mechanisms of infection, transmission, and action of these viral pandemics.

### Reaching across countries

Beyond coordination at a national level, the strength of institutional support for long-term studies rests in their participation in much larger international networks or infrastructures such as the eLTER (Integrated European Long-Term Ecosystem, critical zone and socio-ecological Research) network, the AnaEE (Analysis and Experimentation on Ecosystems) network, or the Global Biodiversity Information Facility (GBIF) network (**Fig. 4**). The inclusion of long-term studies in such networks provide individual researchers with financial, logistic and methodological opportunities to reinforce the intrinsic value of their long-term studies. At the same time, the data generated by long-term studies critically feed into these global initiatives, informing national and international policy. For instance, the novel European Biodiversity Observation Coordination Centre (EBOCC) initiative is designed to establish a lasting framework for coordinating and enhancing biodiversity monitoring across the European Union, its primary goal being to facilitate the production and utilization of high-quality biodiversity data that will serve as a foundation for EU policy decisions on conservation efforts and sustainable ecosystem management. The EBOCC – whose key objectives are (1) data mobilization and integration and (2) improved data collection methods – is built upon national biodiversity monitoring hubs, leveraging existing networks and communities, such as those fostered through institutional support. In this regard, data from SEE-LIFE projects managed at the DoHNEE research infrastructure will directly feed into the national French center for biodiversity data (https://www.pndb.fr/en/home): the French node of the Group on Earth Observations Biodiversity Observation Network[71] (GEO BON; https://geobon.org/bons/national-regional-bon/national-bon/french-bon/).

Since solutions to impacts of global changes on ecosystems and biodiversity cannot emerge without accounting for socioeconomic variables and interactions, global networks are also particularly important in successfully integrating socio-ecological concepts into long-term studies. For instance, many long-term studies of the SEE-LIFE program directly contribute to the long-term socioecological research (LTSER) platforms of the eLTER through the LTSER-France research network ("Zone Ateliers"), consisting of place-based socio-ecological research encompassing key habitats, land uses, and socio-economic units across multiple scales. Such networks not only provide actionable data for stakeholders, they provide a community with the knowledge and tools for successful socio-ecological research, based on what now amounts to decades of experience[72].

### Advocating towards the public and future generations

Last but not least, scientists have an urgent responsibility of *communicating the factual results stemming from long-term* studies to the general public and combating fake-news and misleading information spread by partisan animosity, implicit ideological bias, political polarization, and politically motivated reasoning[73]. Practically, this can take many forms from reports in the grey literature to lay-audience events including forums and conferences, to interviews, books, and movies (for instance, see https://www.youtube.com/watch?v=uWk1Pyk3hhg). Yet, in an increasingly polarized media environment, communicating on scientific results and experiments may be complex, some confidentiality aspects required (for instance when talking about critically endangered habitats or species), or expose individual researchers to backlash from particular groups with interests conflicting with the results produced. Here, the role of the institution (and the endorsement of a consortium from the long-term research community) is to herald researchers' findings through institutions-led communication outlets, with tools and advice from communication and legal teams. For instance, in 2025, the CNRS Ecology & Environment will be coordinating and supporting the launch of a scientific monography series on long-term studies of the SEE-LIFE program with the objective of reaching both the scientific community and stakeholders.  Further still, as partisan-fueled scientific mistrust grows in today's society, re-building a bond of trust between scientists and the public by favoring public investment in long-term research through citizen-science initiatives is urgently needed. Such actions not only serve to educate society at large[74] and advocate for the importance of ecological and evolutionary research, but further allow the collection of actionable data and addressing fundamental questions that would not have been possible through institutionally-led research programs alone[75–77].

# Concluding remarks

The examples provided above showcase why the long-term monitoring of our biosphere is essential for pathing the way into a sustainable future. Only by understanding ecosystems, inhabitants and their responses to change can we hope to enact meaningful policies. We emphasize the importance of institutional support and key levers through which institutions may transcend the impact of individual studies to actionable policies. In recognizing our partners from other countries that have already done so, and following suit in providing institutional support to long-term research programs in ecology and evolution, we hope to encourage a global initiative for supporting the long-term research programs that yield invaluable knowledge on the ecosystems and biodiversity upon which we all depend.

# Acknowledgments

We are sincerely grateful to Véronique Mathet for overseeing all administrative aspects relating to the SEE-Life program; to Floriane Vidal for having overseen all aspects of communication and advertising surrounding the SEE-Life program until 2025, especially the organization of the 1$^{st}$ SEE-Life Conference held in Paris from 9-12 Sept 2024, and the coordination of the SEE-Life booklet: "Comprendre la biodiversité pour agir" (https://www.inee.cnrs.fr/sites/institut_inee/files/download-file/SEE-Life_Livret_BD.pdf), and to Antoine Roux for his continued help on the program. We are particularly grateful to Michel Mortier and Brigitte Roux, general director and deputy director of CNRS Foundation, for their support in raising funds for the SEE-Life program. The SEE-Life program and actions thereof would not be possible if not for the hard-work and dedication of numerous support staff, including Ariane Azario, Laurent Bernadou, Eudora Berniolles, Lise Berthelot, Karine Bessat, Julie Degen, Hadja Diaby, Mathilde Emery, Thomas Jean-Joseph, Isabelle Lendo, Anne Lucas, Isabelle Poulain, Mohamed Said, Linda Salvaneschi and Clarisse Toitot. We sincerely thank them all – and all of those which we failed to name – for working in the shadows in support of long-term scientific research. Finally, our deepest gratitude goes to the incredible, diverse and inclusive community invested in long-term studies – from the insightful founders of these studies through generations of keepers of what today constitutes a truly unique scientific heritage on the natural environment.

# Figures

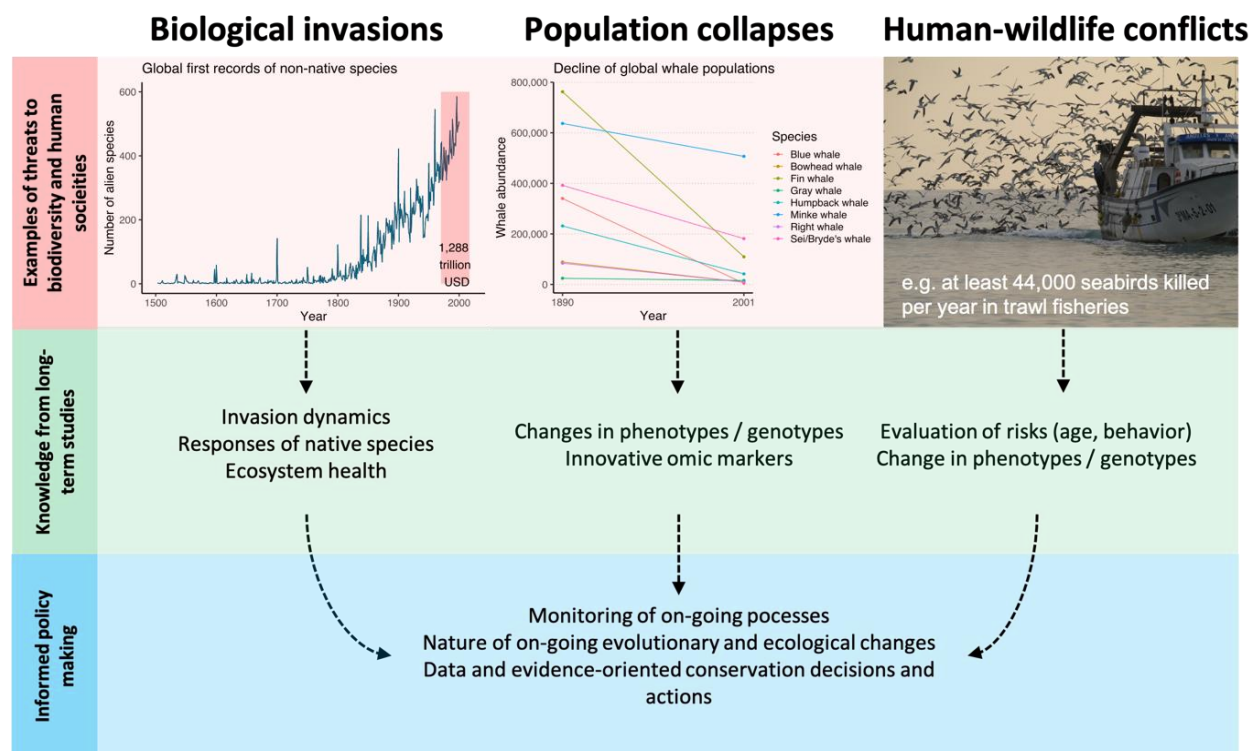

Fig 1. **Examples of known global threats to biodiversity and human societies for which long-term studies in ecology and evolution may shed fundamental insights allowing informed policy actions.** The threats presented include biological invasions, global population collapses of known species, or conflicts between human economic activities and wildlife. For *Biological invasions* and *Population collapses,* the data was obtained online at Our World in Data (https://ourworldindata.org). The photograph used to illustrate *Human-wildlife conflicts* is by Salvador Garcia (IEO) and is reproduced with permission by the author. For *Biological invasions*, the graph shows the number of established alien species and the year they were first recorded out of their native range. A species may be counted multiple times, once for each new country they are found in. Records after 2000 are not shown as data for those years may be underestimated due to a reporting lag. Data includes birds, mammals, reptiles, amphibians, vascular plants, bryophytes, algae, fishes, insects, mollusks, crustaceans, arachnids, arthropods (myriapods, diplopods, etc.), other invertebrates, fungi, bacteria and protozoans. For information purposes the estimate economic cost of biological invasions for the period 1970-2017 is indicated and stems from[24]. For *population collapses*, the graph shows the decline in global whale abundance between the pre-whaling period and 2001. Note that the estimates, especially those from the distant past, come with significant uncertainty (not presented here). For *Human-wildlife conflicts*, the estimate of 44,000 seabirds caught per year in trawl fisheries stems from[78].

Data sources: Seebens et al. (2017)[79] – processed by Our World in Data. "Global first records of non-native species" [dataset]. Seebens et al., "Global Alien Species First Record Database" [original data]. Retrieved April 7, 2025 from

Whale populations (Pershing et al. 2010)[80] – processed by Our World in Data. "Whale population" [dataset]. Whale populations (Pershing et al. 2010) [original data]. Retrieved April 7, 2025 from https://ourworldindata.org/grapher/whale-populations?focus=~Fin+whale

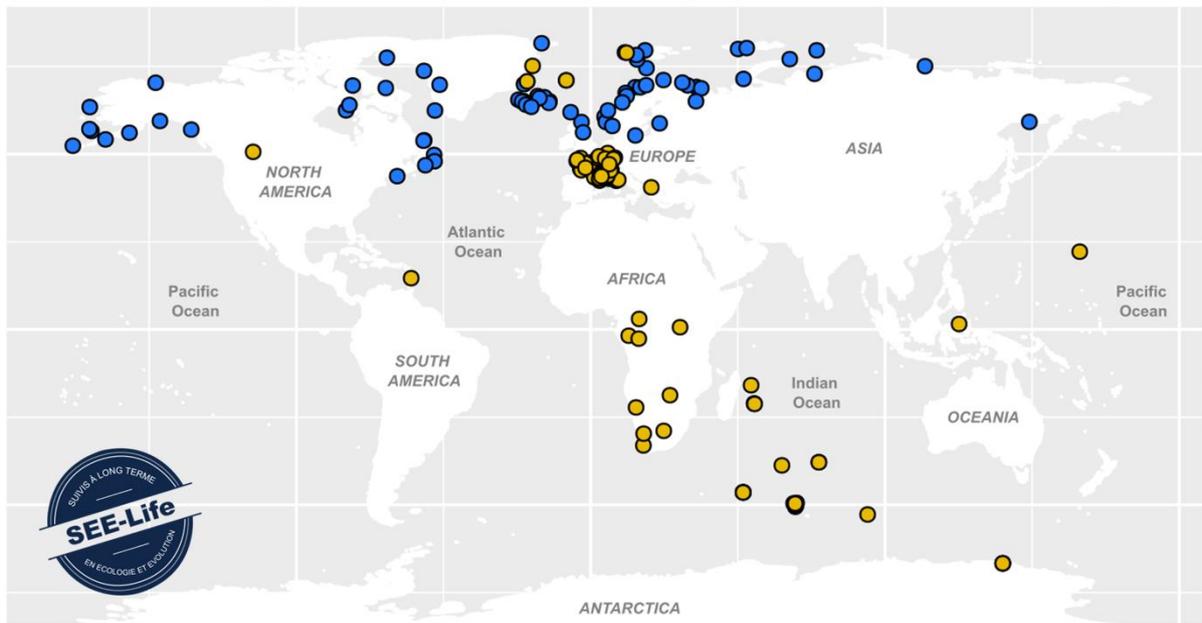

Fig 2. Global map of the long-term Studies in Ecology and Evolution supported in the framework of the SEE-Life program of the French National Center for Scientific Research (CNRS). Map markers in blue in the Arctic belong to the same long-term study focusing on chemical pollution monitoring in Arctic seabirds.

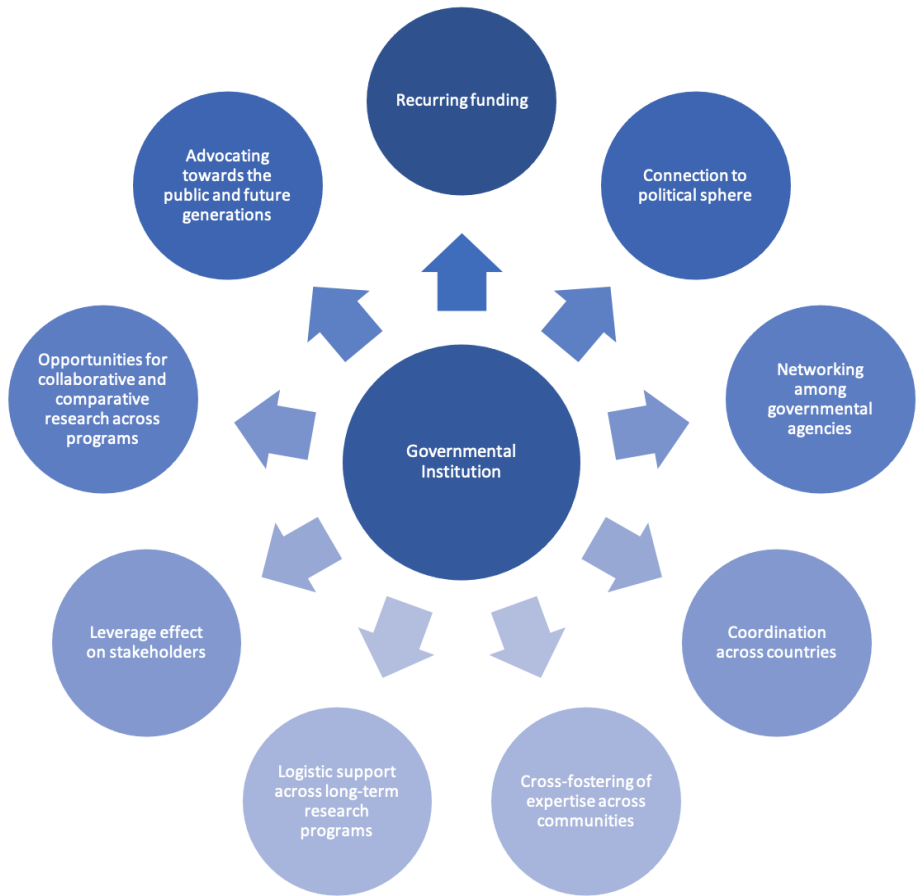

Fig 3. Key areas where institutional support of long-term studies in ecology and evolution may be pivotal in transforming data into informed action.

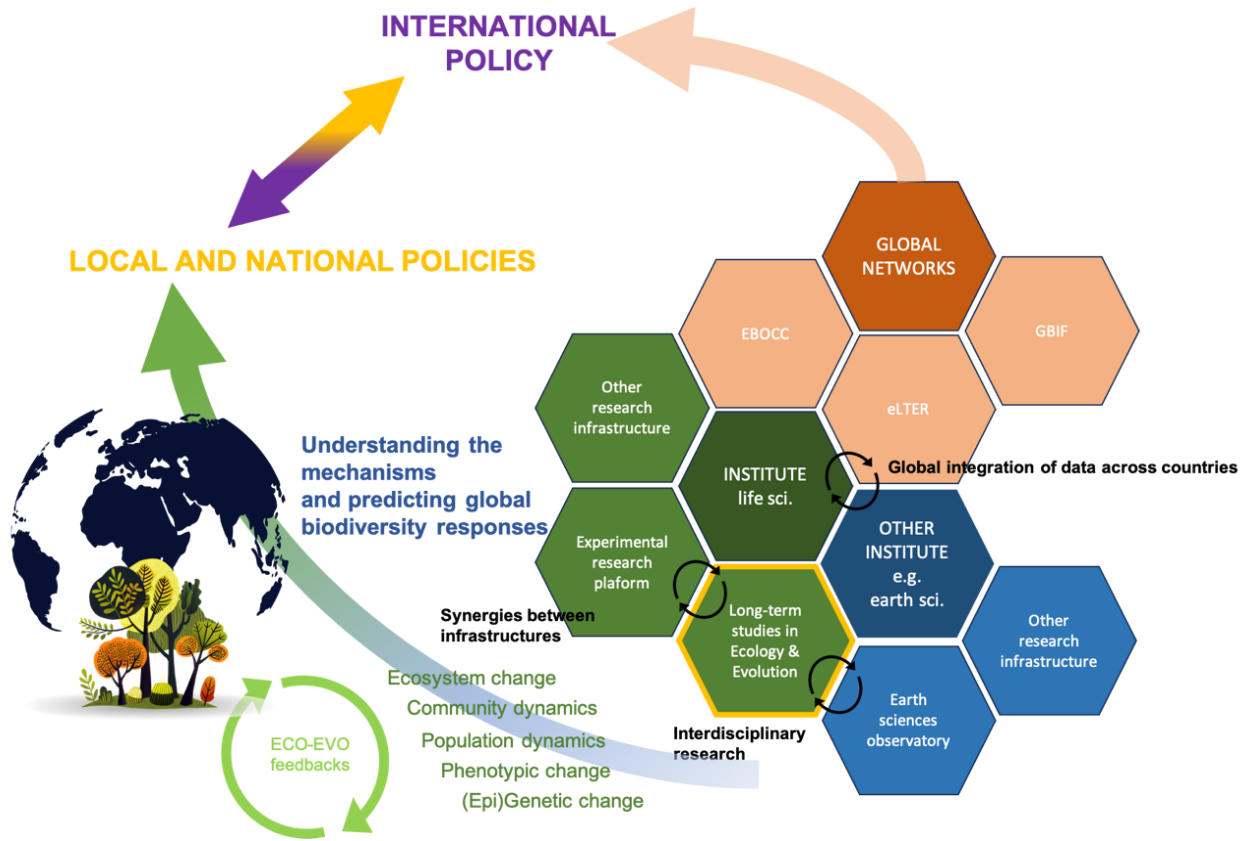

Fig 4. **Long-term studies in ecology and evolution allow key insights into eco-evolutionary feedbacks and the mechanisms underlying organism, community and ecosystem adaptation to global change.** By promoting interdisciplinary research between research departments (e.g., life sciences in green and earth sciences in blue), and networking across national (e.g., other research agencies, not shown) and international spheres (in orange), institutions play a key role in building global data frames allowing to understand and predict global biodiversity responses to change and enact meaningful political action at local, national and international levels. Abbreviations stand for: EBOCC = European Biodiversity Observation Coordination Centre; eLTER = Integrated European Long-Term Ecosystem, critical zone and socio-ecological Research Infrastructure; GBIF = Global Biodiversity Information Facility network.